\documentclass[11pt,letter]{article}
\pdfoutput=1 

\usepackage{natbib}
 \bibpunct{(}{)}{;}{a}{,}{,}
\usepackage{graphicx}
\usepackage{amsmath,amssymb,amsthm}
\usepackage{bm}
\usepackage[vlined]{algorithm2e}
\usepackage{multicol}
\usepackage{titlesec}
\usepackage{latexsym}
\usepackage[pdftex,colorlinks=false,linkcolor=blue,citecolor=blue,urlcolor=blue,bookmarks=true,pdfpagemode=None]{hyperref}
\usepackage{url}
\usepackage{verbatim}
\usepackage{setspace}
\usepackage{fullpage}
\usepackage{titling}
\usepackage{float}

\titlespacing{\section}{0pt}{*1.5}{*1.5}
\titlespacing{\subsection}{0pt}{*1.5}{*1.5}
\titlespacing{\subsubsection}{0pt}{*1.5}{*1.5}

\title{Semi-parametric Robust Event Detection for \\ Massive Time-Domain Databases\protect\thanks{A\lowercase{\rm ddress correspondence to
\href{mailto:ablocker@fas.harvard.edu}{ablocker@fas.harvard.edu}}.
A shorter version of this work appeared in \emph{Statistical Challenges in Modern Astronomy V}, Springer-Verlag, 177--189}}

\author{Alexander W Blocker\thanks{Department of Statistics, Harvard University, \texttt{ablocker@fas.harvard.edu}} \and Pavlos Protopapas\thanks{Harvard-Smithsonian Center for Astrophysics, \texttt{pprotopapas@cfa.harvard.edu}}}

\date{}

\begin{document}

\maketitle

\abstract{
The detection and analysis of events within massive collections of time-series has become an extremely important task for time-domain astronomy.
In particular, many scientific investigations (e.g. the analysis of microlensing and other transients) begin with the detection of isolated events in irregularly-sampled series with both non-linear trends and non-Gaussian noise.
We outline a semi-parametric, robust, parallel method for identifying variability and isolated events at multiple scales in the presence of the above complications.
This approach harnesses the power of Bayesian modeling while maintaining much of the speed and scalability of more ad-hoc machine learning approaches.
We also contrast this work with event detection methods from other fields, highlighting the unique challenges posed by astronomical surveys.
Finally, we present results from the application of this method to 87.2 million EROS-2 sources, where we have obtained a greater than 100-fold reduction in candidates for certain types of phenomena while creating high-quality features for subsequent analyses.
}

\section{Introduction}
\label{sec:intro}

The analysis of massive time-domain astronomical surveys poses growing challenge within astrostatistics that demands both statistical rigor and computational efficiency.
While such data provides a wide range of opportunities, the detection of isolated events is one ubiquitous problem that generally takes on a given outline: We are presented with a massive (10--100+ million) database of time series, possibly spanning multiple spectral bands.
Our goal is to identify and classify time series containing events.
How do we define an event? We are not interested in isolated outliers (as is the case in anomaly detection).
Instead, we are looking for groups of observations that differ significantly from those nearby (i.e. ``bumps'' and ``spikes'').
In our applications of interest, such groups are differentiated from trends by their time scale---that is, they have structure at a higher frequency than we would consider a trend, but with a lower frequency than isolated outliers.
Additionally, we would like to distinguish globally-variable light curves from isolated events, as they have very different scientific interpretations.
This flavor of problem arises in many fields, but the case of astronomical time-domain surveys is particularly challenging.

%

There is an acute need for statistical methods that scale to these volumes of data throughout modern astronomy.
This demands that we carefully manage the trade-off between statistical rigor and computational efficiency.
In general, principled statistical methods yield better performance with messy, complex data, but scale poorly to massive datasets.
In contrast, more ad-hoc machine learning methods handle clean data well, but often choke on issues we confront with complex astronomical data (outliers, nonlinear trends, irregular sampling, unusual dependence structures, etc.).
Our approach is to inject probability modeling into our analysis in the right places, gaining much of the power of probability modeling without incurring its computational penalties.

We demonstrate the utility of this approach using a multi-stage technique for event detection.
By combining a principled, flexible probability model with a discriminative classifier, we obtain excellent performance and computational efficiency analyzing the MACHO and EROS-2 surveys.

\section{Previous Approaches \& Unique Challenges}

The astronomical literature contains a variety of approaches, among which scan statistics are prevalent.
These have seen use in astronomical surveys \citet{LiangEtAl2004,PrestonEtAl2009}, but they often discard information by working with ranks and account for neither trends nor irregular sampling.
Equivalent width methods (a scan statistic based upon local deviations) are also common in astrophysics.
However, these typically rely upon Gaussian assumptions and relatively simple multiple testing corrections; the latter can unnecessarily decrease detection power.
Numerous other approaches have been proposed in the literature, the vast majority of which rely upon Gaussian distributional assumptions, stationary, and/or regular sampling.

This problem also has a long history within the statistical community, often under the moniker of ``change-point'' or ``regime-switching''.
Some recent examples include the work of Smyth and his collaborators \citet{HutchinsEtAl2008,IhlerEtAl2007}, who have used hidden Markov models to model deviations from learned baselines in sensor count data.
There is a strong Bayesian lines of research on this topic; \citet{Smith1975,RafteryAkman1986,CarlinEtAl1992} are representative examples of this work.
On the econometrics side, \citet{Andrews1993} and more recent work by Perron \& collaborators \citet{BaiPerron1998,PerronQu2006} are only a small part of the literature.
Our setting is differs greatly from those seen in the vast majority of previous work.

Most preceding work has dealt with single, long time series which provide a high degree of internal replication.
This allows methods to reliably ascertain what behavior is ``typical'' and find deviations from it with little outside information.
In analyzing massive time-domain surveys, we have large sets of time series that are less informative individually.
We must therefore rely on replication across series and prior scientific knowledge to find deviations from ``typical'' behavior.
Furthermore, we must handle the additional complications of astronomical data.

These complications arise from both the measurement processes used in astronomical studies and the nature of the phenomena we study.
The distribution of measurement errors from ground-based observations is typically heavy-tailed (extreme outliers are prevalent).
The resulting data requires more sophisticated noise models than the typical Gaussian.
Non-linear, low-frequency trends are also common due to long-period variation in source intensity and/or  calibration.
Such trends render naive, trend-free methods less effective; in particular, their specificity diminishes in this setting.
The related but distinct problem of non-event light curves with variation at the time scale of interest also complicates our analysis and demands tools that can discriminate between these cases.
Finally, irregular sampling is ubiquitous in astronomical surveys due to changes in the earth's orientation throughout the year and other factors.
Irregular sampling can create artificial events in analyses that discard observation times; therefore, our method must take this information into account to maintain both high specificity and high sensitivity.

\section{Models \& Methods}
\label{sec:model}

Our analysis consists of two stages.
First, we use a Bayesian probability model to detect of sources with variation at a time scale of interest (i.e. the time scale of events) and to reduce the dimensionality of our time series (using posterior summaries).
Second, we employ a classifier based on these posterior summaries to discriminate among different types of variability.
In the application described in Sections \ref{sec:data} and \ref{sec:results}, these types correspond to periodic and temporally-isolated (event-like) variability.

Formally, let $V$ be the set of all time series with variation at a given time scale of interest (e.g., the range of lengths for isolated events), and let $S$ be a subset of $V$ corresponding to the time series of interest (events).
For a given light curve $Y_i$, we want to estimate $P(Y_i \in S)$; that is, the probability that it is an event.

We decompose this probability as
\begin{equation}
P(Y_i \in S) = P( Y_i \in V \cap S ) = P(Y_i \in V) \cdot P(Y_i \in S | Y_i \in
V)
\end{equation}
estimating or bounding each probability separately using the techniques described above.
This decomposition allows us to employ generative techniques in the first stage while harnessing discriminative techniques in the second.
We provide details of the models underlying these techniques below and cover the corresponding inference algorithms in Section \ref{sec:computation}.

\subsection{Semi-parametric model for variable light curves}
\label{subsec:waveletmodel}

To flexibly model both non-linear trends and events at the time-scale of interest, we turn to wavelets.
Their localization in both time and frequency allows us to separate event-like variation (characterized by a higher frequency) from trends (characterized by a lower frequency) while preserving local structure of our light curves.

We begin by specifying a linear model for each time series with a ``split'' incomplete wavelet basis:
\begin{equation}
y(t) = \beta_0 \phi_0(t) + \sum_{i=1}^{k_l} \beta_i \phi_i(t) +
 \sum_{j=k_l+1}^{M} \beta_j \phi_j(t) + \epsilon(t)
\end{equation}
Here, $y(t)$ is the observed magnitude at time $t$.
We define $(\phi_1, \ldots, \phi_{k_l})$ as the $k_l$ lowest-frequency components of a discrete-frequency wavelet basis, and $(\phi_{k_l+1}, \ldots, \phi_{M})$ as the higher-frequency components.
The idea is for $(\phi_1, \ldots, \phi_{k_l})$ to model structure due to trends, and $(\phi_{k_l+1}, \ldots, \phi_{M})$ to model structure at the scales of interest for events.
We use an incomplete basis (excluding the highest frequencies) as we are not interested in modeling variation at time scales below those of interest for our events.

This basis formulation explicitly addresses irregular sampling as well.
We simply evaluate the basis functions at the observation times to obtain a valid model for our light curve.
This is simpler and more adaptable than, for example, using a continuous time autoregressive model.

To stabilize our inferences and regularize our estimates in under-sampled time periods (gaps), we impose a $N(0, \sigma^2 / \tau)$ prior on $(\beta_1, \ldots, \beta_M)$.
This is conditionally conjugate to an augmented form of our model, which allows for efficient inference.
The prior parameter $\tau$ is also readily interpretable: it is the number of artificial observations we are introducing for each coefficient.
We set $\tau = \frac{1}{100}$ for our inference to reflect a diffuse prior; it is, however, sufficient to regularize our estimates in under-sampled periods.

To account for the extreme outliers observed in our light curves, we assume that our residuals $\epsilon(t)$ are distributed as independent $t_{\nu}(0,\sigma^2)$ random variables.
This allows our inference to ignore isolated outliers, focusing on variation with more structure.
We fix $\nu$ for our model at $5$; it is possible, although computationally expensive, to infer $\nu$ as well.

Selection of the wavelet basis $\phi$ is an important consideration for this method.
It determines the trade-off between time and frequency localization for our inference, and it also constrains (due to incompleteness) the types of variation we can approximate well.
In general, this choice depends upon the scientific context.
We select the Symmlet 4 (a.k.a. Least Asymmetric Daubechies 4) wavelet basis for this work for its high degree of time localization, reasonable frequency localization, and quality of approximation for the phenomena of interest.

The final remaining choices are interval over which the basis is defined (to which our observation times are rescaled), the dimensionality of our basis $M$, and the number of ``trend'' components $k_l$.
All three of these are interrelated and must be selected based on the time-scale of interest for events (as opposed to trends).
We scale our basis to an interval of length $2048$ and set $M=128$, $k_l=8$.
This provides enough resolution to capture events at the scale of interest while removing low-frequency trends and isolated outliers.

\subsection{Screening for variation at frequencies of interest}
\label{subsec:screening}

We screen light curves for further examination by testing $H_0: \, \beta_{k_l+1} = \beta_{k_l+2} = \ldots = \beta_M = 0$ against the alternative that any of these coefficients differs from zero.
This procedure will select many light curves that do not contain isolated events, but its primary purpose is to provide a set of candidate light curves of manageable size for further investigation and classification.
Selected non-event light curves contain variation at the scale of interest, but this variation may be temporally diffuse.
Our test statistic is $2(\hat{\ell}_1 - \hat{\ell}_0)$, where $\hat{\ell}_0$ is the log-likelihood of the null model evaluated at the MAP estimates; $\hat{\ell}_1$ is the analogous quantity for the alternative model.
We use a $\chi^2$ approximation for the reference distribution of this test statistic.
Although this approximation is technically incorrect given the use of an informative prior, it provides a reasonable approximation that holds empirically.
With this approximation, we employ a modified Benjamini-Hochberg FDR procedure with a maximum FDR of $10^{-4}$ to set the critical region for our test statistic \citet{BenjaminiHochberg1995,BenjaminiYekutieli2001}.
We present our validation for this technique in Section \ref{subsec:results-screening}.

\subsection{Classification model for isolated variation}
\label{subsec:discriminative}

We engineered two features based on the model in Section \ref{subsec:waveletmodel} to discriminate between diffuse and isolated variability in the light curves selected by our screening procedure.
Both are based on the normalized output of the preceding model, as this allows us to remove the nonlinear trends and isolated outliers.
We thus obtain a high-quality, detrended and denoised representation of each light curve.
We define for each light curve
\begin{equation}
 \tilde{y}(t) = \sum_{j=k_l+1}^{M} \hat{\beta}_j \phi_j(t) \;;\quad
 z(t) = \frac{\tilde{y}(t) -
\mathrm{Mean}(\tilde{y}(t))}{\mathrm{SD}(\tilde{y}(t))}
\end{equation}

Our first feature is a monotonic transformation of a conventional CUSUM statistic, defined as $CUSUM$ via
\begin{equation}
 S(t) = \sum_{s \leq t} (z(s)^2 - 1) \;;\quad
 CUSUM = \log( 1 + \frac{\max_t S(t) - \min_t S(t)}{\sqrt{n}} )
\end{equation}
This captures the degree of temporal concentration for the variation in our fitted values---larger values will correspond to localized deviations from the baseline, while low values will correspond to deviations spread over a greater duration.
It is maximized for a single spike with a flat baseline.

Our second feature is ``directed variation''.
Our goal is for it to capture deviation from symmetric variation (as would be observed in periodic or quasi-periodic light curves).
Letting $z_{\mathrm{med}}$ be the median of $z(t)$, we define:
\begin{eqnarray}
DV &=& \frac{1}{\#\{t : z(t) > z_{\mathrm{med}}\}}
 \sum_{t : z(t) > z_{\mathrm{med}}} z(t)^2 -
 \frac{1}{\#\{t : z(t) < z_{\mathrm{med}}\}}
 \sum_{t : z(t) < z_{\mathrm{med}}} z(t)^2
\end{eqnarray}

We tested a variety of classifiers including SVM (with linear and radial kernels), $k$NN, and LDA.
However, in the end, we obtained our best performance from regularized logistic regression.
We used a ``weakly informative'' prior as developed by \citet{GelmanEtAl2008} to stabilize the estimates from this model.
We describe its training and evaluation in Section \ref{subsec:discriminative_training}.

\section{Computation}
\label{sec:computation}

Speed and scalability are the core goals of our computational strategy.
We require a method that scales to databases of 200 million or more light curves (for the EROS-2 survey).
As a result, our inference is optimization-based (as opposed to simulation) and highly-tuned for efficiency.
We also manage the scale of training data where possible, preventing the computational cost of inference from scaling poorly with database size. We lay out the particulars of our algorithms below.
A C implementation of the EM algorithm described in Section \ref{subsec:em_inference}, an R implementation of the screening procedure described in Section \ref{subsec:screening}, and an R script to construct the features described in \ref{subsec:discriminative} are available under the LGPL v2.1 license in the \texttt{rowavedt} package via \url{https://www.github.com/awblocker/rowavedt}.

\subsection{Efficient EM inference for semi-parametric model}
\label{subsec:em_inference}

To obtain estimates of $\beta_0, \ldots, \beta_M$ and $\sigma^2$ in our semi-parametric model, we first augment our model with a set of observation-specific variances.
Let $z(t) \sim N(0,1)$ independent of $w(t) \sim \mathrm{InvGamma}(\frac{\nu}{2},\frac{\nu}{2})$.
Then, we can represent $\epsilon(t)$ as $\epsilon(t) \sim z(t) \cdot \sqrt{w(t)}$.
This allows us to consider the set of $w(t)$ as missing data, opening our model to tools such as the EM algorithm \citet{DempsterLairdRubin1977}.

Following this approach, we employ an EM algorithm with the optimal data augmentation scheme of \citet{MengVanDyk1997} to obtain MAP estimates for the parameters of our semi-parametric model.
Compared to a naive EM implementation, we have found that this scheme offers a 5 to 10-fold reduction in the number of iterations required for convergence.

We implemented this procedure in C with a direct interface to an optimized BLAS/LAPACK implementation\footnote{We used both ATLAS and Intel MKL; the latter provided a 20--30\% speedup over the former.}.
Small numerical details in this implementation had a dramatic effect on our  final computational efficiency.
In particular, directly solving the normal equations via a Cholesky decomposition for the regression in our M-step provided a 7 to 8-fold speedup over using the QR decomposition (as is standard).
Because we must update our weights (and hence all matrix products in our regression) at each iteration of the EM algorithm, such gains have a major impact on our final run-times.

This allowed us to obtain an average time per complete estimation procedure (including EM estimation for both the null and alternative models, as specified in Section \ref{subsec:screening}) of approximately 0.15--0.2 seconds, including file I/O, using a single processor on Harvard's Odyssey cluster.
Memory usage was below 16MB per light curve, and this algorithm is embarrassingly parallel across light curves.
This combination allows our technique to scale to extremely large sets of time series.

\subsection{Training the classification model via simulation}
\label{subsec:discriminative_training}

We train our classification model on a combination of simulated data and curated, labeled light curves.
Before descending into the details, we emphasize that this model must distinguish between local and global variation in light curves that have already passed the first-stage screen.
Thus, our training data includes only such light curves.

\begin{figure}[H]
\centering
 \includegraphics[width=0.5\textwidth]{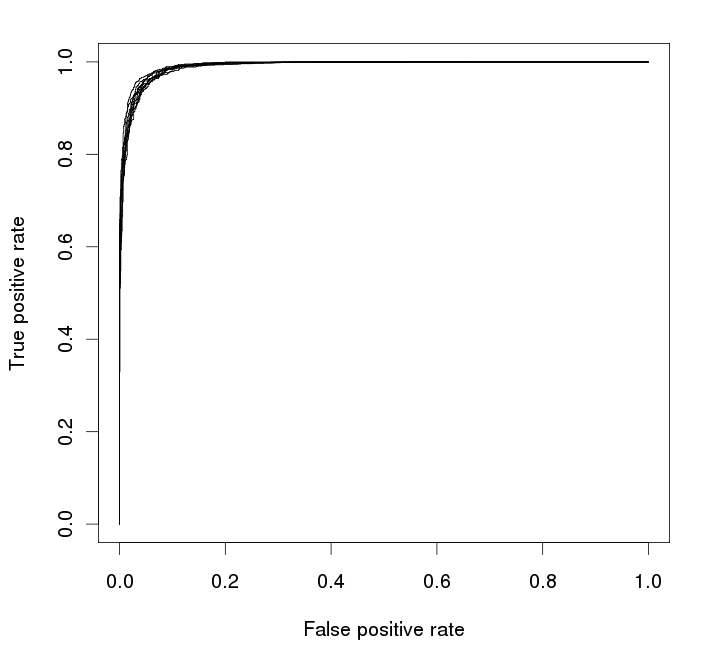}
 \caption{ROCs from 10-fold cross-validation on MACHO training data}
 \label{fig:ROC}
\end{figure}

The training data consisted of 12,365 labeled variable light curves from the MACHO dataset (periodic and quasi-periodic) and 9,170 simulated events (microlensing) that passed the given screening procedure.
We obtained \emph{maximum a posteriori} (MAP) estimates for the parameters of this model via numerical maximization and performed 10-fold cross-validation to assess its predictive ability.
This validation showed excellent performance, with a mean cross-validated AUC of 0.991 on our training data.
The ROC curves are shown in Figure \ref{fig:ROC}.

\section{Data}
\label{sec:data}

We used used data from the MACHO survey for training and testing.
The knowledge and information gained from this data was then used to analyze the EROS-2 survey.

The MACHO database consists of approximately 38 million LMC (Large Magellanic Cloud) sources, each observed in two spectral bands \citet{AlcockEtAl1993,BennettEtAl1996,AlcockEtAl2001}.
Data was collected from 1992 through 1999 on 50-inch telescope at Mount Stromlo Observatory, Australia on 94 43’x 43’ fields in two bands, using eight 2048 x 2048 pixel CCD’s.
This data contains substantial gaps in observations due to seasonality and competing priorities for transient events.

The EROS-2 database consists of approximately 87.2 million sources, each observed in two spectral bands.
Imaging was conducted with a 1m telescope at ESO, La Silla between 1996 and 2003, each camera consisting of mosaic of eight 2K x 2K LORAL CCDs.
There are typically 800--1000 observations per source, and outlying observations are prevalent (although less so than in the MACHO data).

\section{Results}
\label{sec:results}

Here we present our results on the MACHO and EROS-2 surveys.
We begin by examining the behavior of our semi-parametric model and its estimation procedure (as described in Sections \ref{subsec:waveletmodel} and \ref{subsec:em_inference}).
We then turn to our frequency-based screening method (described in Section \ref{subsec:screening}), focusing on its operating characteristics and performance on EROS-2 data.
Finally, we examine the behavior of our classifier (described in Sections \ref{subsec:discriminative} and \ref{subsec:discriminative_training}), considering the distributions of our features and the qualitative properties of light curves classified as events.

\subsection{Semi-parametric model --- empirical properties}
\label{subsec:results-model}

\begin{figure}[H]
 \begin{center}
 \includegraphics[width=0.45\textwidth]{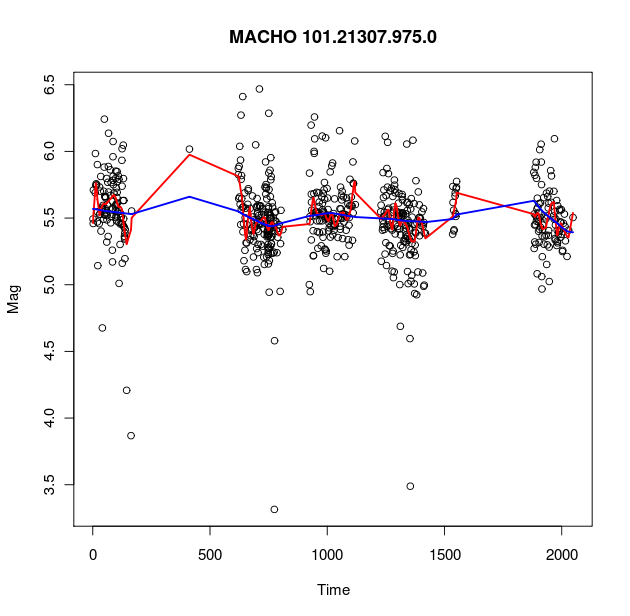}
 \includegraphics[width=0.45\textwidth]{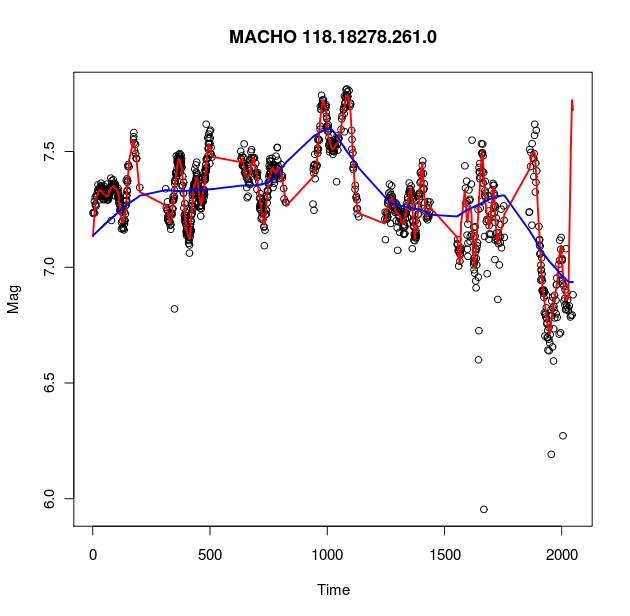}
 \includegraphics[width=0.45\textwidth]{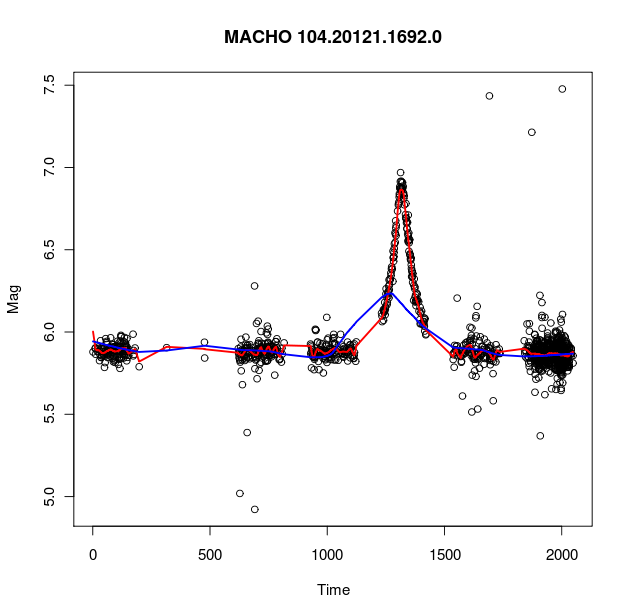}
 \caption{Examples of fits for null , variable, and event MACHO light
curves (clockwise from top). Null model is in blue; complete model is in red.}
 \label{fig:fits}
 \end{center}
\end{figure}

The semi-parametric model provided reasonable fits for both MACHO and EROS-2 data.
It captured both non-linear trends (including changes in baseline between observing periods).
We provide examples of fits for both the null and complete model on null and event light curves in Figure \ref{fig:fits}.

\subsection{Screening}
\label{subsec:results-screening}

\begin{figure}[H]
\centering
 \includegraphics[width=0.4\textwidth]{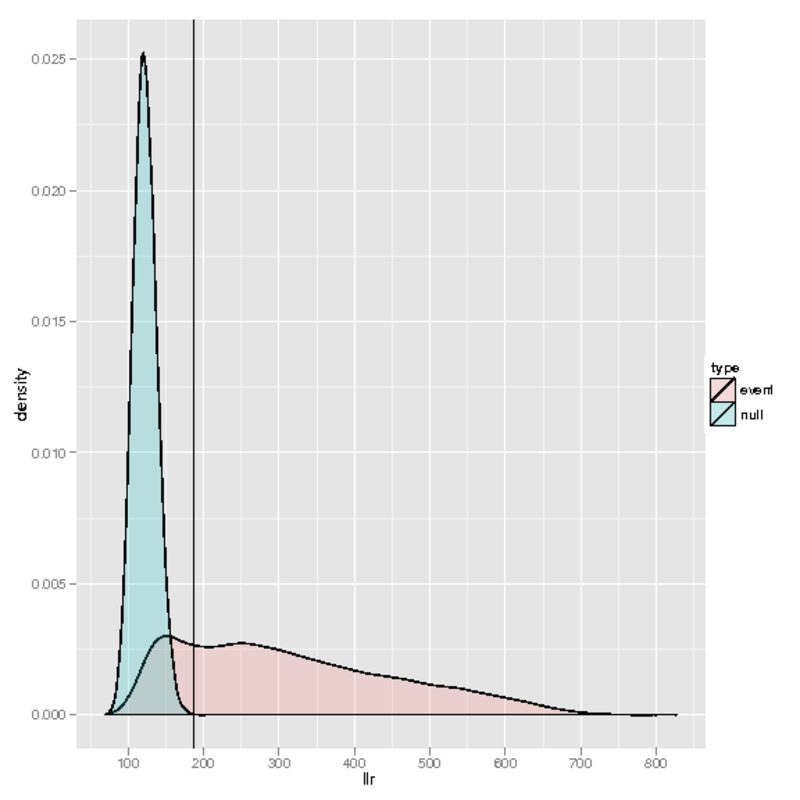}
 \caption{Comparative distribution of LLR statistic for simulated null and
 event light curves}
 \label{fig:llr-sim}
\end{figure}

To assess how well our LLR statistic and overall screening procedure performs on the data of immediate interest, we simulated $50,000$ events from a physics-based model (for microlensing) and $50,000$ null time series based on the observed properties of the MACHO data.
We obtain approximate power of $80\%$ with an FDR of $10^{-4}$ based on this simulated data.
We include a visualization of the resulting critical value for our LLR statistic and the separation of these distributions in Figure \ref{fig:llr-sim}.
 
Running this on the EROS-2 data, we obtain a reduction of approximately 98\% (from 87.2 million candidate light curves to approximately 1.5 million) from our screening procedure.
This greatly eased the computational burden of subsequent analyses.

\subsection{Classification of isolated events}
\label{subsec:results-classification}

\begin{figure}[H]
 \centering
 \includegraphics[height=0.455\textheight]{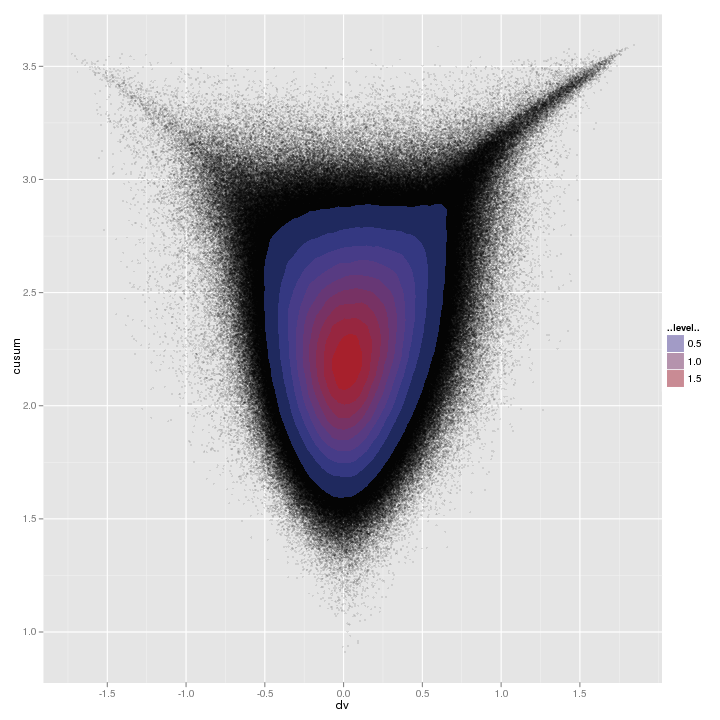}

 \includegraphics[height=0.455\textheight]{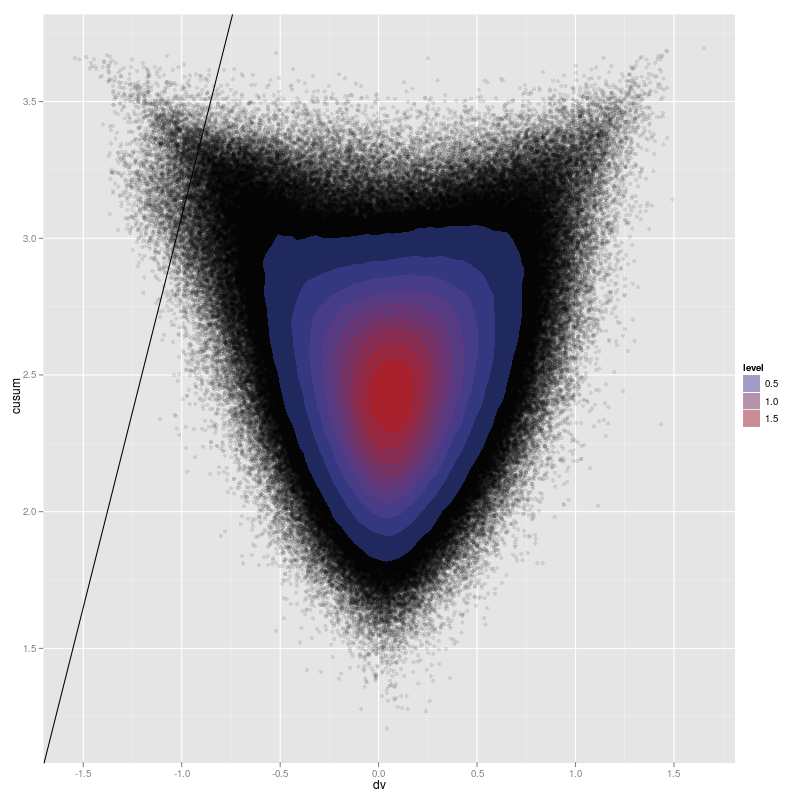}
 \caption{Distribution of classification features for MACHO (top) and EROS-2 (bottom) databases. $DV$ on the horizontal axis, $CUSUM$ on the vertical axis. Classification boundary from logistic regression shown for EROS-2 data.}
 \label{fig:feature-dists}
\end{figure}

Our classifier selected approximately 49,000 of the screened light curves as likely isolated events ($P \geq 0.5$).
Of these, approximately 17,000 survived a final round of screening before further investigation.
This final screen consisted of removing all fields with 20 or more identified events, as such clusters were not of scientific interest for the current investigation.
One major example of this from EROS-2 is the supernova SN1987a, which affected light curves from the Large Magellanic Cloud.
For other investigations, however, such screening may not be appropriate or necessary.
We show the distribution of features for MACHO and EROS-2, with the estimated classification boundary, in Figure \ref{fig:feature-dists}.

\begin{figure}[H]
 \includegraphics[width=0.5\textwidth, trim=0 0 0 24pt, clip=true]{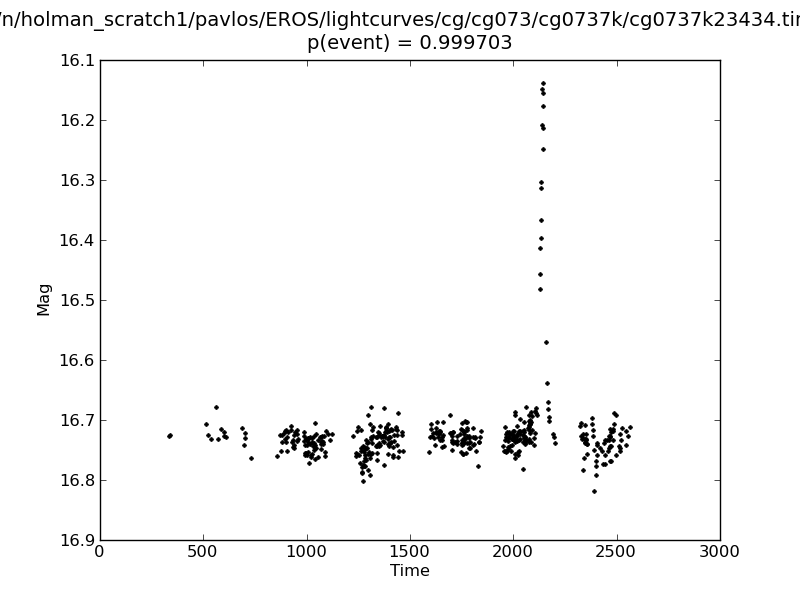}
 \includegraphics[width=0.5\textwidth, trim=0 0 0 24pt, clip=true]{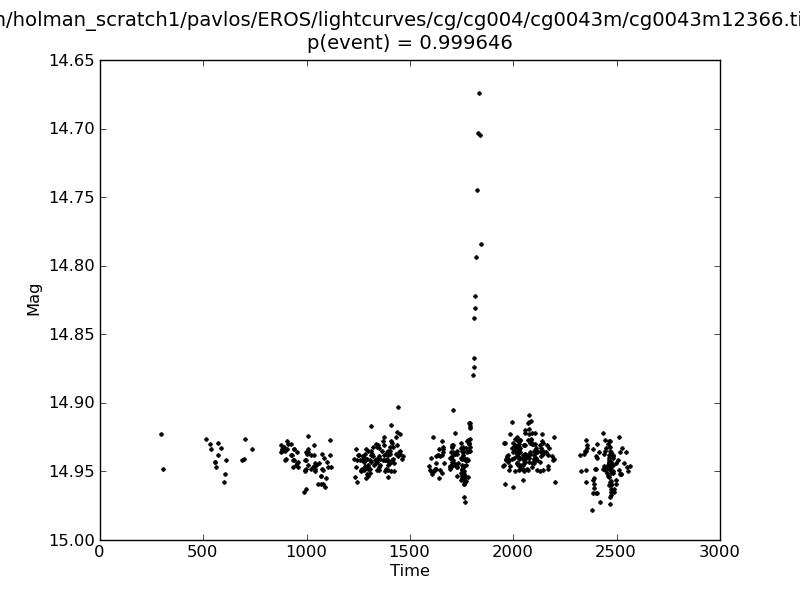}
 \includegraphics[width=0.5\textwidth, trim=0 0 0 24pt, clip=true]{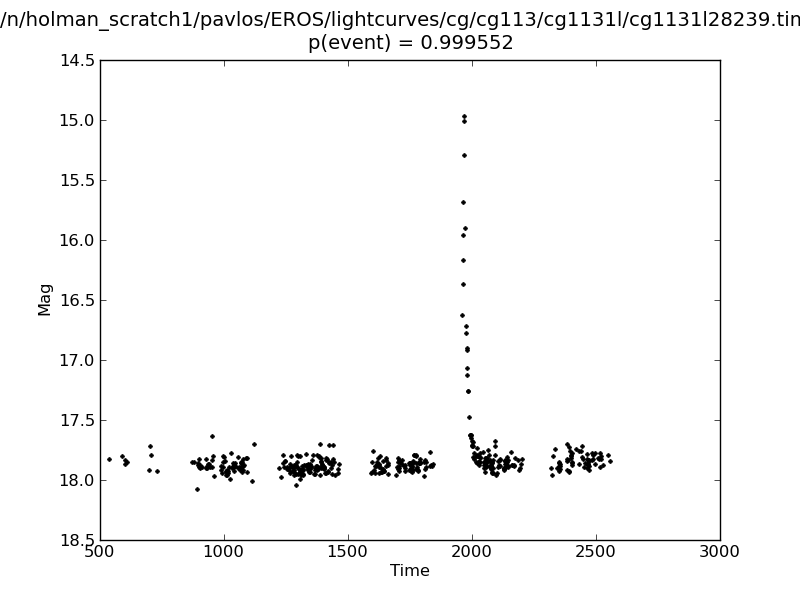}
 \includegraphics[width=0.5\textwidth, trim=0 0 0 24pt, clip=true]{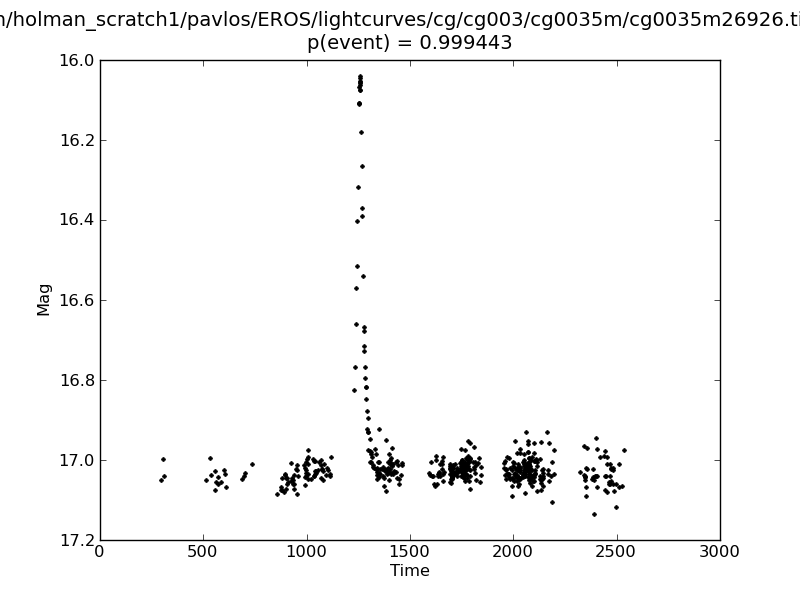}
 \caption{Top four detected events from the EROS-2 database}
 \label{fig:detected}
\end{figure}

Within the events detected for EROS-2, we have found 68 known microlensing events, 42 known supernovas, and 25 known Cepheids with an (admittedly incomplete) database search (VizieR only).
We have also identified several hundred previously unidentified transient phenomena that we are investigating further.
These have been validated as previously unlabeled against a thorough database search (VizieR, Simbad, and VO).
We provide plots of the top four detected events in Figure \ref{fig:detected}.

\section{Remarks}
\label{sec:remarks}

The method we have demonstrated combines the power of principled probability modeling with the speed and flexibility of more ad-hoc machine learning approaches.
It scales to the analysis of massive astronomical time-domain surveys and can be adapted to detect a variety of temporally-isolated phenomena.
It does not provide a final, scientific classification or analysis for light curves in these surveys; rather, we want to predict which time series are most likely to yield phenomena characterized by events (e.g. microlensing, blue stars, flares, etc.).
Our technique is, at its core, a tool for rigorously-grounded discovery rather than approximate final analysis.

This, in turn, allows for the use of more complex, physically-motivated model on massive databases by pruning the set relevant data to a manageable size.
We accomplish this while providing assessments of uncertainties at each stage of our screening and detection, and we provide a sufficiently rich framework to incorporate relevant domain knowledge.

We look forward to the application of this technique to more surveys and phenomena; in particular, we are currently investigating data from Pan-STARRS.
The approach demonstrated here can be applied to many other massive data challenges within astronomy and beyond, bringing the power of Bayesian probability modeling to massive data while maintaining computational tractability.

\section*{Acknowledgements}
We would like to thank Dae-Won Kim, Dan Preston, and Jean-Baptiste Marquette for their assistance with the MACHO and EROS-2 datasets.
We would also like to thank Xiao-Li Meng, Edoardo Airoldi, David van Dyk, Aneta Siemiginowska, and Vinay L. Kashyap for their feedback and discussions.

\bibliographystyle{chicago}
\bibliography{references}

\begin{thebibliography}{}

\bibitem[\protect\citeauthoryear{{Alcock}, {Allsman}, {Alves}, {Axelrod},
  {Becker}, {Bennett}, {Cook}, {Drake}, {Freeman}, {Geha}, {Griest}, {Lehner},
  {Marshall}, {Minniti}, {Nelson}, {Peterson}, {Popowski}, {Pratt}, {Quinn},
  {Stubbs}, {Sutherland}, {Tomaney}, {Vandehei}, and {Welch}}{{Alcock}
  et~al.}{2001}]{AlcockEtAl2001}
{Alcock}, C., R.~A. {Allsman}, D.~R. {Alves}, T.~S. {Axelrod}, A.~C. {Becker},
  D.~P. {Bennett}, K.~H. {Cook}, A.~J. {Drake}, K.~C. {Freeman}, M.~{Geha},
  K.~{Griest}, M.~J. {Lehner}, S.~L. {Marshall}, D.~{Minniti}, C.~A. {Nelson},
  B.~A. {Peterson}, P.~{Popowski}, M.~R. {Pratt}, P.~J. {Quinn}, C.~W.
  {Stubbs}, W.~{Sutherland}, A.~B. {Tomaney}, T.~{Vandehei}, and D.~{Welch}
  (2001, October).
\newblock {The MACHO Project: Microlensing Detection Efficiency}.
\newblock {\em The Astrophysical Journal Supplement Series\/}~{\em 136},
  439--462.

\bibitem[\protect\citeauthoryear{{Alcock}, {Allsman}, {Axelrod}, {Bennett},
  {Cook}, {Park}, {Marshall}, {Stubbs}, {Griest}, {Perlmutter}, {Sutherland},
  {Freeman}, {Peterson}, {Quinn}, and {Rodgers}}{{Alcock}
  et~al.}{1993}]{AlcockEtAl1993}
{Alcock}, C., R.~A. {Allsman}, T.~S. {Axelrod}, D.~P. {Bennett}, K.~H. {Cook},
  H.~S. {Park}, S.~L. {Marshall}, C.~W. {Stubbs}, K.~{Griest}, S.~{Perlmutter},
  W.~{Sutherland}, K.~C. {Freeman}, B.~A. {Peterson}, P.~J. {Quinn}, and A.~W.
  {Rodgers} (1993, January).
\newblock {The MACHO Project - a Search for the Dark Matter in the Milky-Way}.
\newblock In {B.~T.~Soifer} (Ed.), {\em Sky Surveys. Protostars to
  Protogalaxies}, Volume~43 of {\em Astronomical Society of the Pacific
  Conference Series}, pp.\  291--+.

\bibitem[\protect\citeauthoryear{Andrews}{Andrews}{1993}]{Andrews1993}
Andrews, D. W.~K. (1993).
\newblock Tests for parameter instability and structural change with unknown
  change point.
\newblock {\em Econometrica\/}~{\em 61\/}(4), pp. 821--856.

\bibitem[\protect\citeauthoryear{Bai and Perron}{Bai and
  Perron}{1998}]{BaiPerron1998}
Bai, J. and P.~Perron (1998).
\newblock Estimating and testing linear models with multiple structural
  changes.
\newblock {\em Econometrica\/}~{\em 66\/}(1), pp. 47--78.

\bibitem[\protect\citeauthoryear{Benjamini and Hochberg}{Benjamini and
  Hochberg}{1995}]{BenjaminiHochberg1995}
Benjamini, Y. and Y.~Hochberg (1995).
\newblock Controlling the false discovery rate: A practical and powerful
  approach to multiple testing.
\newblock {\em Journal of the Royal Statistical Society. Series B
  (Methodological)\/}~{\em 57\/}(1), pp. 289--300.

\bibitem[\protect\citeauthoryear{Benjamini and Yekutieli}{Benjamini and
  Yekutieli}{2001}]{BenjaminiYekutieli2001}
Benjamini, Y. and D.~Yekutieli (2001).
\newblock The control of the false discovery rate in multiple testing under
  dependency.
\newblock {\em The Annals of Statistics\/}~{\em 29\/}(4), pp. 1165--1188.

\bibitem[\protect\citeauthoryear{{Bennett}, {Alcock}, {Allsman}, {Axelrod},
  {Cook}, {Freeman}, {Griest}, {Marshall}, {Peterson}, {Pratt}, {Quinn},
  {Rodgers}, {Stubbs}, and {Sutherland}}{{Bennett}
  et~al.}{1996}]{BennettEtAl1996}
{Bennett}, D.~P., C.~{Alcock}, R.~A. {Allsman}, T.~S. {Axelrod}, K.~B. {Cook},
  K.~C. {Freeman}, K.~{Griest}, S.~L. {Marshall}, B.~A. {Peterson}, M.~R.
  {Pratt}, P.~J. {Quinn}, A.~W. {Rodgers}, C.~W. {Stubbs}, and W.~{Sutherland}
  (1996).
\newblock {The MACHO Project Dark Matter Search}.
\newblock In {V.~Trimble \& A.~Reisenegger} (Ed.), {\em Clusters, Lensing, and
  the Future of the Universe}, Volume~88 of {\em Astronomical Society of the
  Pacific Conference Series}, pp.\  95--+.

\bibitem[\protect\citeauthoryear{Carlin, Gelfand, and Smith}{Carlin
  et~al.}{1992}]{CarlinEtAl1992}
Carlin, B.~P., A.~E. Gelfand, and A.~F.~M. Smith (1992).
\newblock Hierarchical bayesian analysis of changepoint problems.
\newblock {\em Journal of the Royal Statistical Society. Series C (Applied
  Statistics)\/}~{\em 41\/}(2), pp. 389--405.

\bibitem[\protect\citeauthoryear{Dempster, Laird, and Rubin}{Dempster
  et~al.}{1977}]{DempsterLairdRubin1977}
Dempster, A.~P., N.~M. Laird, and D.~B. Rubin (1977).
\newblock Maximum likelihood from incomplete data via the em algorithm.
\newblock {\em Journal of the Royal Statistical Society. Series B
  (Methodological)\/}~{\em 39\/}(1), pp. 1--38.

\bibitem[\protect\citeauthoryear{Gelman, Jakulin, Pittau, and Su}{Gelman
  et~al.}{2008}]{GelmanEtAl2008}
Gelman, A., A.~Jakulin, M.~G. Pittau, and Y.~Su (2008).
\newblock A weakly informative default prior distribution for logistic and
  other regression models.
\newblock {\em The Annals of Applied Statistics\/}~{\em 2\/}(4), 1360--1383.

\bibitem[\protect\citeauthoryear{Hutchins, Ihler, and Smyth}{Hutchins
  et~al.}{2008}]{HutchinsEtAl2008}
Hutchins, J., A.~Ihler, and P.~Smyth (2008).
\newblock {Probabilistic analysis of a large-scale urban traffic data set}.
\newblock In {\em Proceedings of the Second International Workshop on Knowledge
  Discovery from Sensor Data (ACM SIGKDD Conference, KDD-08}. Citeseer.

\bibitem[\protect\citeauthoryear{Ihler, Hutchins, and Smyth}{Ihler
  et~al.}{2007}]{IhlerEtAl2007}
Ihler, A., J.~Hutchins, and P.~Smyth (2007).
\newblock {Learning to detect events with Markov-modulated poisson processes}.
\newblock {\em ACM Transactions on Knowledge Discovery from Data (TKDD)\/}~{\em
  1\/}(3), 13--es.

\bibitem[\protect\citeauthoryear{Liang, Rice, Pater, Alcock, Axelrod, Wang, and
  Marshall}{Liang et~al.}{2004}]{LiangEtAl2004}
Liang, C.-L., J.~A. Rice, I.~d. Pater, C.~Alcock, T.~Axelrod, A.~Wang, and
  S.~Marshall (2004).
\newblock Statistical methods for detecting stellar occultations by kuiper belt
  objects: The taiwanese-american occultation survey.
\newblock {\em Statistical Science\/}~{\em 19\/}(2), pp. 265--274.

\bibitem[\protect\citeauthoryear{Meng and Dyk}{Meng and
  Dyk}{1997}]{MengVanDyk1997}
Meng, X.-L. and D.~v. Dyk (1997).
\newblock The em algorithm--an old folk-song sung to a fast new tune.
\newblock {\em Journal of the Royal Statistical Society. Series B
  (Methodological)\/}~{\em 59\/}(3), pp. 511--567.

\bibitem[\protect\citeauthoryear{Perron and Qu}{Perron and
  Qu}{2006}]{PerronQu2006}
Perron, P. and Z.~Qu (2006).
\newblock Estimating restricted structural change models.
\newblock {\em Journal of Econometrics\/}~{\em 134\/}(2), 373--399.

\bibitem[\protect\citeauthoryear{{Preston}, {Protopapas}, and
  {Brodley}}{{Preston} et~al.}{2009}]{PrestonEtAl2009}
{Preston}, D., P.~{Protopapas}, and C.~{Brodley} (2009, January).
\newblock {Event Discovery in Time Series}.
\newblock {\em ArXiv e-prints. To appear in SIAM International Conference on
  Data Mining\/}~{\em arxiv:0901.3329v1 [astro-ph.IM]}.

\bibitem[\protect\citeauthoryear{Raftery and Akman}{Raftery and
  Akman}{1986}]{RafteryAkman1986}
Raftery, A.~E. and V.~E. Akman (1986).
\newblock Bayesian analysis of a poisson process with a change-point.
\newblock {\em Biometrika\/}~{\em 73\/}(1), pp. 85--89.

\bibitem[\protect\citeauthoryear{Smith}{Smith}{1975}]{Smith1975}
Smith, A. F.~M. (1975).
\newblock A bayesian approach to inference about a change-point in a sequence
  of random variables.
\newblock {\em Biometrika\/}~{\em 62\/}(2), 407--416.

\end{thebibliography}

\end{document}